\documentclass[lettersize,journal]{IEEEtran}
\usepackage{amsmath,amsfonts,amssymb,amsthm}
\usepackage{xcolor}
\usepackage{algorithm}
\usepackage{algpseudocode}
\usepackage{array}
\usepackage{textcomp}
\usepackage{stfloats}
\usepackage{url}
\usepackage{verbatim}
\usepackage{graphicx}
\usepackage{cite}
\usepackage{multirow}
\usepackage{caption}
\usepackage{subcaption}
\usepackage{hhline}

\def\BibTeX{{\rm B\kern-.05em{\sc i\kern-.025em b}\kern-.08em
    T\kern-.1667em\lower.7ex\hbox{E}\kern-.125emX}}
\usepackage{balance}

  \def\cC{{\mathcal{C}}}

 \def\cN{{\mathcal{N}}}  
   
   \def\cX{{\mathcal{X}}}
\def\cY{{\mathcal{Y}}}

\def\bPhi{{\pmb{\Phi}}} 
 \def\bgamma{{\pmb{\gamma}}}

\def\b0{{\pmb{0}}} 

\def\ba{{\mathbf{a}}}   
   \def\bh{{\mathbf{h}}}
   
 \def\bn{{\mathbf{n}}}  
  \def\bs{{\mathbf{s}}} 
   
\def\by{{\mathbf{y}}}   

\def\bA{{\mathbf{A}}}   
\def\bE{{\mathbf{E}}}   \def\bH{{\mathbf{H}}}
\def\bI{{\mathbf{I}}}   
\def\bM{{\mathbf{M}}} \def\bN{{\mathbf{N}}}  
  \def\bS{{\mathbf{S}}} 
   
\def\bY{{\mathbf{Y}}}

\begin{document}
	
\title{Downlink Channel Estimation for mmWave Systems with Impulsive Interference}

\author{\thanks{This work was supported by Samsung Electronics Co., Ltd (IO250508-12767-01); by Institute of Information \& communications Technology Planning \& Evaluation (IITP) grant funded by the Korea government (MSIT) (No. RS-2024-00395824, Development of Cloud virtualized RAN (vRAN) system supporting upper-midband); and by Institute of Information \& communications Technology Planning \& Evaluation (IITP) grant funded by the Korea government (MSIT) (No.2021-000269, Development of sub-THz band wireless transmission and access core technology for 6G Tbps data rate).}%
Kwonyeol Park,\thanks{K. Park is with the S.LSI Division, Samsung Electronics Company Ltd., Hwaseong-si, 18448, South Korea, and also with the School of Electrical Engineering, Korea Advanced Institute of Science and Technology, Daejeon 34141, South Korea (e-mail: kwon10.park@kaist.ac.kr).}
Gyoseung Lee\thanks{G. Lee and J. Choi are with the School of Electrical Engineering, Korea Advanced Institute of Science and Technology, Daejeon 34141, South Korea (e-mail: \{iee4432, junil\}@kaist.ac.kr).}, Hyeongtaek Lee\thanks{H. Lee is with the Department of Electronic and Electrical Engineering, Ewha Womans University, Seoul 03760, South Korea (e-mail: htlee@ewha.ac.kr).}, 
Hwanjin Kim\thanks{H. Kim is with the School of Electronics Engineering, Kyungpook National University, Daegu 41566, South Korea (e-mail: hwanjin@knu.ac.kr).}, and Junil Choi
}
\maketitle

\begin{abstract}
In this paper, we investigate a channel estimation problem in a downlink millimeter-wave (mmWave) multiple-input multiple-output (MIMO) system, which suffers from impulsive interference caused by hardware non-idealities or external disruptions. Specifically, impulsive interference presents a significant challenge to channel estimation due to its sporadic, unpredictable, and high-power nature. To tackle this issue, we develop a Bayesian channel estimation technique based on variational inference (VI) that leverages the sparsity of the mmWave channel in the angular domain and the intermittent nature of impulsive interference to minimize channel estimation errors. The proposed technique employs mean-field approximation to approximate posterior inference and integrates VI into the sparse Bayesian learning (SBL) framework. Simulation results demonstrate that the proposed technique outperforms baselines in terms of channel estimation accuracy. 
\end{abstract}


\section{Introduction}
\label{Intro}

Impulsive interference, which happens due to internal hardware impairments or external electromagnetic disturbances, can be a performance-limiting factor in wireless communication systems.
Unlike additive white Gaussian noise (AWGN), impulsive interference is sporadic and manifests as sudden high-power spikes. This characteristic makes it particularly disruptive to channel estimation in millimeter-wave (mmWave) systems, where accurate channel estimation is demanding \cite{kim2023bayesian}. In practical systems, especially on the user equipment (UE) side, cost-efficient hardware designs often rely on low-cost components and increased component sharing to reduce manufacturing expenses. However, these design choices introduce additional impairments, such as non-linear power amplifiers, I/Q imbalance, and quantization errors, which not only degrade signal quality but also amplify the impact of impulsive interference \cite{epple2016advanced}. As a result, conventional channel estimation techniques based on the AWGN assumption struggle to maintain reliable performance in the presence of impulsive interference, which highlights the need for more robust estimation methods.

A variety of studies have focused on mitigating the impact of impulsive interference. However, most of these techniques assume that impulsive interference is already known \cite{alam2020mitigation}, which is unrealistic. Recent studies have introduced joint estimation approaches for both channel and impulsive interference, leveraging compressed sensing and Bayesian algorithms \cite{zhou2020uplink}. In particular, Bayesian inference has been widely used to estimate unknown impulsive interference, providing a probabilistic framework that incorporates prior knowledge and uncertainty. However, these studies have primarily employed Student-$t$ models to characterize impulsive interference \cite{shang2015robust, guo2023variational}. Unfortunately, this model may not fully capture its sharpness and bursty nature, necessitating a more appropriate distribution. 

In this work, we propose a robust downlink channel estimation algorithm for mmWave systems under impulsive interference. To effectively capture the non-Gaussian nature of impulsive interference, we adopt a sparse Bayesian learning (SBL) framework for inference and model the interference using the complex adaptive Laplace distribution \cite{bai2023space}. Unlike conventional techniques, our approach leverages the sparsity-promoting characteristics of the complex adaptive Laplace distribution, making it well-suited for modeling impulsive interference \cite{BaBacan:2010}. However, incorporating such a sparsity-inducing prior into Bayesian inference leads to analytical intractability, making it challenging to compute the exact posterior distribution. To address this, we employ variational inference (VI) to efficiently approximate the posterior distribution, enabling joint estimation of the channel and impulsive interference components. Simulation results demonstrate that the proposed technique achieves superior channel estimation performance, thereby validating the effectiveness of the tailored prior design.

The rest of this paper is organized as follows. In Section~II, we introduce the system model and formulate the problem of interest. In Section~III, we describe the proposed VI-based channel estimation algorithm, which is built upon the SBL framework. Section~IV provides simulation results and performance analysis. Finally, Section~V concludes the paper.

\section{System Model and Problem Formulation}
\label{sec2}
In this section, we first introduce the system and channel models and then formulate the problem of estimating the downlink channel and impulsive interference.
\subsection{System and channel models}
\label{sec2.A}
We consider a downlink mmWave multiple-input multiple-output orthogonal frequency division multiplexing (MIMO-OFDM) system, where a gNodeB (gNB) with $N_{\mathrm{t}}$ antennas in a uniform linear array (ULA) serves a single UE with $N_{\mathrm{r}}$ antennas in the ULA. In OFDM systems, each subcarrier experiences flat-fading, allowing channel estimation to be performed independently for each subcarrier. The channel is assumed to be quasi-static during the estimation period. The downlink received signal for a given subcarrier at the $t$-th time slot is expressed as
\begin{align} \label{y_t}
    \mathbf{y}[t] &= \bar{\mathbf{H}}\mathbf{s}[t] + \mathbf{e}[t] + \mathbf{n}[t],
\end{align} 
where $\bar{\bH} \in \mathbb{C}^{N_{\mathrm{r}} \times N_{\mathrm{t}}}$ is the downlink channel from the gNB to the UE, $\bs[t]=[s_1[t],\cdots,s_{N_{\mathrm{t}}}[t]]^{\mathrm{T}} \in \mathbb{C}^{N_{\mathrm{t}} \times 1}$ is the pilot signal transmitted from the gNB to the UE at the $t$-th time slot satisfying $\mathbb{E}[\vert s_k[t]\vert^2]=P$ with gNB transmit power $N_{\mathrm{t}}P$, and $\bn[t] \sim \cC\cN(\b0_{N_\mathrm{r}}, \sigma^2 \bI_{N_\mathrm{r}})$ is additive white Gaussian noise (AWGN) with variance $\sigma^2$.
Here, $\mathbf{e}[t]$ represents an unknown impulsive interference, which occurs sporadically with abrupt high-power spikes.\footnote{While standardized mitigation approaches exist for multiplicative hardware impairments like phase noise through dedicated reference signals \cite{3gpp.38.211, choi2023phase}, additive impulsive interference poses unique challenges due to its sporadic nature.} We model $\mathbf{e}[t]$ such that most elements are zero, with non-zero values occurring at random spatiotemporal locations, without imposing specific distributional assumptions since the proposed technique is generally applicable. Based on this modeling, stacking (\ref{y_t}) during $T$ time slots, the observation matrix at the UE $\bY \in \mathbb{C}^{N_{\mathrm{r}} \times T}$ is given by
\begin{align}\label{Y}
    \mathbf{Y} &= [\by[1],\cdots,\by[T]] \nonumber \\ &= \bar{\mathbf{H}}\mathbf{S} + \mathbf{E} + \mathbf{N},
\end{align} 
where $\bS=[\bs[1],\cdots,\bs[T]] \in \mathbb{C}^{N_{\mathrm t} \times T}$, $\bE = [\mathbf{e}[1],\cdots,\mathbf{e}[T]] \in \mathbb{C}^{N_{\mathrm r} \times T}$, and $\bN = [\bn[1],\cdots,\bn[T]] \in \mathbb{C}^{N_{\mathrm r} \times T}$.

As a channel model, we adopt the geometric channel model as in \cite{choi2023withray} that properly describes limited scattering environments in mmWave bands. In this model, the gNB-UE channel $\bar{\bH}$ is expressed as
\begin{align}\label{eq:geometric channel}
    \bar{\bH} &= \sqrt{\frac{N_{\mathrm r}N_{\mathrm t}}{L}} \sum_{\ell=1}^{L} \alpha_{\ell}\ba_{N_{\mathrm r}}\left(\theta_{\ell}\right)\ba_{N_{\mathrm t}}^{\mathrm{H}}\left(\phi_{\ell}\right),
\end{align}
where $L$ is the number of propagation paths in the gNB-UE channel, and $\alpha_{\ell}$ is the complex channel gain of the $\ell$-th path, which depends on the path loss. The $N_{\mathrm{r}}$-dimensional ULA array steering vector at the UE $\ba_{N_{\mathrm{r}}}\left(\cdot\right)$ is
\begin{align} \label{a_N}
    \mathbf{a}_{N_{\mathrm{r}}}(\theta_{\ell}) = \frac{1}{\sqrt{N_{\mathrm{r}}}}[1,\ e^{j\frac{2\pi d_{\mathrm{U}}}{\lambda}\cos(\theta_{\ell})},\ \cdots\ ,e^{j\frac{2\pi d_{\mathrm{U}}({N_{\mathrm{r}}}-1)}{\lambda}\cos(\theta_{ \ell})}]^{\mathrm{T}},
\end{align}
where $\lambda$ and $d_{\mathrm{U}}$ are the carrier wavelength and the antenna spacing at the UE, respectively. The array steering vector $\ba_{N_{\mathrm{t}}}\left(\cdot\right)$ at the gNB is similarly defined as in (\ref{a_N}) with the gNB antenna spacing $d_{\mathrm{B}}$.

\subsection{Problem formulation}
In mmWave systems, signals typically propagate through a limited number of path clusters, making the angular domain channels highly sparse. The following virtual angular domain transformation allows the use of the angular domain sparsity present in the gNB-UE channel:
\begin{align} \label{H_virtual}
\bar{\bH}=\mathbf{A}_\mathrm{U}\mathbf{H}\mathbf{A}^{\mathrm{H}}_\mathrm{B},
\end{align}
where $\mathbf{A}_\mathrm{U} \in \mathbb{C}^{N_{\mathrm{r}} \times D_{\mathrm{U}}}$ and $\mathbf{A}_\mathrm{B} \in \mathbb{C}^{N_{\mathrm{t}} \times D_{\mathrm{B}}}$ denote the overcomplete dictionaries for the UE and gNB, satisfying $D_{\mathrm{U}} \geq N_{\mathrm{r}}$ and $D_{\mathrm{B}} \geq N_{\mathrm{t}}$, respectively, and $\mathbf{H} \in \mathbb{C}^{D_{\mathrm{U}} \times D_{\mathrm{B}}}$ is the equivalent angular domain gNB-UE channel. The dictionary for the UE $\mathbf{A}_\mathrm{U}$ is
\begin{align}
\bA_{\mathrm{U}} =[\ba_{N_{\mathrm{r}}}(\hat{\theta}_1),\cdots,\ba_{N_{\mathrm{r}}}(\hat{\theta}_{D_{\mathrm{U}}})],
\end{align}
where $\{\hat{\theta}_1, \cdots, \hat{\theta}_{D_{\mathrm{U}}} \}$ are predefined grids, and the dictionary for the BS $\mathbf{A}_\mathrm{B}$ is similarly defined.\footnote{By enabling soft energy allocation, our proposed approach offers robustness to modeling uncertainties, such as grid mismatch.}
Based on (\ref{H_virtual}), the observation matrix at the UE $\bY$ in (\ref{Y}) can be rewritten as 
\begin{align}\label{eq:received signal}
    \mathbf{Y} &= \mathbf{A}_\mathrm{U}\mathbf{H}\mathbf{A}^{\mathrm{H}}_\mathrm{B}\mathbf{S} + \mathbf{E} + \mathbf{N}.
\end{align} 

From (\ref{eq:received signal}), our goal is to derive the minimum mean squared error (MMSE) estimates of $\{ \bH, \bE\}$ from $\bY$. Note that the burst nature of impulsive interference induces sparsity in the matrix $\bE$, and thus we can apply the SBL framework to estimate both $\bH$ and $\bE$. Our proposed technique, which will be presented in Section \ref{sec3}, utilizes the VI approach to derive the posterior distributions of $\{ \bH, \bE\}$.

\section{Proposed VI-Based Downlink Channel Estimation Under SBL Framework} \label{sec3}
In this section, we establish a hierarchical Bayesian model that incorporates the sparsity of the channel and impulsive interference. Then, we derive a variational distribution to estimate the posterior distributions of the unknown parameters. Finally, we present the update rules for the proposed technique.
\subsection{Hierarchical Bayesian model} \label{sec3.A}
Using the identity $\mathrm{vec}(\bM_1 \bM_2 \bM_3) = (\bM_3^{\mathrm{T}} \otimes \bM_1)\mathrm{vec}(\bM_2)$, where $\mathrm{vec}(\cdot)$ denotes the vectorization operation, the  measurement matrix $\bY$ in (\ref{eq:received signal}) can be vectorized~as
\begin{align}\label{eq:vectorized signal}
    \mathrm{vec}(\bY) &= \underbrace{(\bS^{\mathrm{T}} \bA_{\mathrm{B}}^{*} \otimes \bA_{\mathrm{U}})}_{=\bPhi} \underbrace{\mathrm{vec}(\bH)}_{=\bh} + \underbrace{\mathrm{vec}(\bE)}_{=\boldsymbol{\mathrm{e}}} + \underbrace{\mathrm{vec}(\bN)}_{=\boldsymbol{\mathrm{n}}} \nonumber \\ & \triangleq \by.
\end{align}
Based on (\ref{eq:vectorized signal}), the conditional distribution of $\boldsymbol{\mathrm{y}}$ is 
\begin{align}\label{eq:cond_y}
p(\by \vert \bh, \boldsymbol{\mathrm{e}}, \beta)&= \mathcal{CN}(\boldsymbol{\Phi} \bh+\boldsymbol{\mathrm{e}}, \beta^{-1}\boldsymbol{\mathrm{I}}_{N_{\mathrm{r}}T}),
\end{align}
where $\beta$ corresponds to the inverse of noise variance at the UE that is generally unknown in practice, modeled as the gamma distribution given by
\begin{align} \label{eq:beta}
p( \beta)&= \mathrm{Gamma} \left(\beta \vert a, b\right) \nonumber \\ &= \frac{b^a}{\Gamma(a)}\beta^{a-1}\exp(-b\beta),
\end{align}
where $\Gamma\left(\cdot\right)$ denotes the gamma function, and $a$ and $b$ are respectively the shape and rate parameters. Note that the Gaussian-gamma distribution has a conjugate-prior relationship, which facilitates tractable posterior inference \cite{hoff2009first}.

In our proposed technique, we model the prior for the channel $\bh$ using a Student-$t$ distribution, which is widely used to effectively capture the angular domain sparsity in mmWave channels \cite{shang2015robust}. Specifically, introducing a random variable $\boldsymbol{\lambda}_{\boldsymbol{\mathrm{h}}}$, the conditional distribution of $\bh$ is
\begin{align} \label{eq:cond_h}
p(\boldsymbol{\mathrm{h}} \vert \boldsymbol{\lambda}_{\boldsymbol{\mathrm{h}}})&= \mathcal{CN}(\boldsymbol{0}_{D_{\mathrm{U}}D_{\mathrm{B}}}, \boldsymbol{\Lambda}_{\boldsymbol{\mathrm{h}}}^{-1}),
\end{align}
where $\boldsymbol{\Lambda}_{\boldsymbol{\mathrm{h}}} = \mathrm{diag}(\boldsymbol{\lambda}_{\boldsymbol{\mathrm{h}}})$, and $\boldsymbol{\lambda}_{\boldsymbol{\mathrm{h}}}$ is modeled as the gamma distribution given by
\begin{align} \label{eq:cond_lambh}
p(\boldsymbol{\lambda}_{\boldsymbol{\mathrm{h}}})&= \prod_{i=1}^{D_{\mathrm{U}}D_{\mathrm{B}}} \mathrm{Gamma} (\lambda_{\boldsymbol{\mathrm{h}}, i}\vert a,b).
\end{align}

As discussed in Section \ref{sec2.A}, to model the characteristics of impulsive interference whose probability density is highly concentrated and sharp, we adopt a complex adaptive Laplace prior\footnote{Through the numerical analysis, we verified that the use of the complex adaptive Laplace distribution closely approximates the true distribution of impulsive interference.} as in \cite{bai2023space}, a complex-adaptive version of the Laplace distribution, constructed by the two-layer hierarchical model by introducing random variables $\{ \boldsymbol{\lambda}_{\boldsymbol{\mathrm{e}}}, \bgamma \}$ as follows:
\begin{align} \label{eq:cond_e}
p(\boldsymbol{\mathrm{e}} \vert \boldsymbol{\lambda_\mathrm{e}})&= \mathcal{CN}(\boldsymbol{0}_{N_{\mathrm{r}}T}, \boldsymbol{\Lambda_\mathrm{e}}),
\\ p(\boldsymbol{\lambda}_{\boldsymbol{\mathrm{e}}} \vert \boldsymbol{\gamma} )&= \prod_{j=1}^{N_{\mathrm{r}}T} \mathrm{Gamma} \left(\lambda_{\boldsymbol{\mathrm{e}}, j} \middle\vert \frac{3}{2},\frac{\gamma_j}{4}\right), \label{eq:cond_lambe} 
\\ p( \boldsymbol{\gamma}) &= \prod_{j=1}^{N_{\mathrm{r}}T} \mathrm{Gamma} \left(\gamma_{j} \vert a, b\right), \label{eq:gamma}
\end{align}
where $\boldsymbol{\Lambda}_{\boldsymbol{\mathrm{e}}} = \mathrm{diag}(\boldsymbol{\lambda}_{\boldsymbol{\mathrm{e}}})$. Note that the marginal distribution of $p(\boldsymbol{\mathrm{e}} \vert \boldsymbol{\gamma})$ corresponds to the complex adaptive Laplace prior. Although this marginal prior is not a conjugate prior to the Gaussian likelihood, the hierarchical construction ensures that each conditional distribution remains conjugate \cite{bai2023space}.
\subsection{VI approach under SBL} \label{sec3.B}
Based on (\ref{eq:cond_y})-(\ref{eq:gamma}), we aim to infer the posterior probability of $\mathcal{X} = \left\{\boldsymbol{\mathrm{h}}, \boldsymbol{\mathrm{e}}, \boldsymbol{\mathrm{\lambda_h}}, \boldsymbol{\mathrm{\lambda_e}}, \boldsymbol{\gamma}, \beta \right\}$ given the observation $\mathcal{Y} = \left\{\by \right\}$, which is generally intractable. To tackle this issue, we use the VI approach under the SBL framework to obtain an approximate posterior distribution. The goal of VI is to derive a variational distribution $q(\cX)$ as close as possible to the true posterior
$p\left( \mathcal{X} \vert \mathcal{Y} \right)$, which is mathematically equivalent to minimizing the Kullback-Leibler (KL) divergence between $q(\cX)$ and $p\left( \mathcal{X} \vert \mathcal{Y} \right)$. Since deriving $q(\cX)$ directly is also challenging, we adopt the mean-field approximation for analytical tractability that enables the following decomposition: $q\left(\mathcal{X}\right)=\prod_m q\left(\mathcal{X}_m\right)$, where $\mathcal{X}_m$ denotes an element of $\mathcal{X}$ \cite{tzikas2008variational}.  
Under this assumption, the optimality condition for $q(\cX)$ of the considered problem can be expressed as  
\begin{align}\label{eq:q_x_i}
q(\cX_m) = \exp \left \{ \langle \log p(\cX, \cY) \rangle_{-\cX_m} \right \}, \enspace \forall m,
\end{align}
where $\langle \cdot \rangle_{ -\mathcal{X}_m}$ stands for the expectation over $\prod_{n \neq m} q\left( \mathcal{X}_n \right)$. In the following, for notational simplicity, we will use $\langle \cdot \rangle$ as the expectation with respect to all variables in $\cX$.
Based on \eqref{eq:q_x_i}, our objective is to determine the functional form of $q\left(\mathcal{X}_m \right), \forall m$ and compute their posterior means, from which the MMSE estimates of $\bh$ and $\mathbf{e}$ can be obtained.
\subsection{Detailed update rules} \label{sec3.C}
Now, we derive the functional forms of $q\left(\mathcal{X}_m\right)$ to obtain an approximate posterior distribution in $\mathcal{X} = \left\{\boldsymbol{\mathrm{h}}, \boldsymbol{\mathrm{e}}, \boldsymbol{\mathrm{\lambda_h}}, \boldsymbol{\mathrm{\lambda_e}}, \boldsymbol{\gamma}, \beta \right\}$.

\textit{1) Derivation of $q\left( \boldsymbol{\mathrm{h}}\right)$ and $q\left( \boldsymbol{\mathrm{e}}\right)$}: First, plugging in \eqref{eq:cond_y} and \eqref{eq:cond_h} to \eqref{eq:q_x_i}, $q(\bh)$ is computed as
\begin{align}\label{eq:q_h}
q\left(\boldsymbol{\mathrm{h}}\right) \propto &\exp \left\{ \langle \log p(\by \vert \boldsymbol{\mathrm{h}}, \boldsymbol{\mathrm{e}}, \beta) + \log p(\boldsymbol{\mathrm{h}} \vert \boldsymbol{\lambda}_{\boldsymbol{\mathrm{h}}}) \rangle_{-\boldsymbol{\mathrm{h}}} \right\} \notag \\
\propto &\exp \left\{ -\left(\boldsymbol{\mathrm{h}} - \boldsymbol{\mu}_{\boldsymbol{\mathrm{h}}} \right)^{\mathrm{H}} \boldsymbol{\Sigma}_{\boldsymbol{\mathrm{h}}}^{-1} \left(\boldsymbol{\mathrm{h}} - \boldsymbol{\mu}_{\boldsymbol{\mathrm{h}}} \right)  \right\}, 
\end{align}
which is a Gaussian distribution with the mean $\boldsymbol{\mu}_{\boldsymbol{\mathrm{h}}}$ and the covariance $\boldsymbol{\Sigma}_{\boldsymbol{\mathrm{h}}}$ respectively given by
\begin{align} \label{mean_h}
\boldsymbol{\mu}_{\boldsymbol{\mathrm{h}}} &= \langle\beta\rangle \boldsymbol{\Sigma}_{\boldsymbol{\mathrm{h}}} \boldsymbol{\Phi}^{\mathrm{H}}\left( \by -\langle \boldsymbol{\mathrm{e}} \rangle \right), 
\boldsymbol{\Sigma}_{\boldsymbol{\mathrm{h}}} = \left( \langle\beta\rangle \boldsymbol{\Phi}^{\mathrm{H}} \boldsymbol{\Phi} + \langle\boldsymbol{\Lambda_{\mathrm{h}}} \rangle\right)^{-1}.
\end{align}
Similarly, $q\left( \boldsymbol{\mathrm{e}}\right)$ is derived by substituting \eqref{eq:cond_y} and \eqref{eq:cond_e} into \eqref{eq:q_x_i}: 
\begin{align}\label{eq:q_e}
q\left(\boldsymbol{\mathrm{e}}\right) \propto &\exp \left\{ \langle \log p(\by \vert \boldsymbol{\mathrm{h}}, \boldsymbol{\mathrm{e}}, \beta) + \log p(\boldsymbol{\mathrm{e}} \vert \boldsymbol{\lambda}_{\boldsymbol{\mathrm{e}}}) \rangle_{-\boldsymbol{\mathrm{e}}} \right\} \notag \\
\propto &\exp \left\{ -\left(\boldsymbol{\mathrm{e}} - \boldsymbol{\mu}_{\boldsymbol{\mathrm{e}}} \right)^{\mathrm{H}} \boldsymbol{\Sigma}_{\boldsymbol{\mathrm{e}}}^{-1} \left(\boldsymbol{\mathrm{e}} - \boldsymbol{\mu}_{\boldsymbol{\mathrm{e}}} \right)  \right\}, 
\end{align}
which implies that $q(\mathbf{e})$ follows a Gaussian distribution associated with the mean $\boldsymbol{\mu}_{\boldsymbol{\mathrm{e}}}$ and the covariance $\boldsymbol{\Sigma}_{\boldsymbol{\mathrm{e}}}$ given by
\begin{align} \label{mean_e}
\boldsymbol{\mu}_{\boldsymbol{\mathrm{e}}} &= \langle\beta\rangle \boldsymbol{\Sigma}_{\boldsymbol{\mathrm{e}}} \left( \by -\boldsymbol{\Phi}\langle  \boldsymbol{h} \rangle \right), 
\boldsymbol{\Sigma}_{\boldsymbol{\mathrm{e}}} = \left( \langle\beta\rangle \cdot \boldsymbol{\mathrm{I}}_{N_{\mathrm{r}}T} + \langle\boldsymbol{\Lambda}^{-1}_{\boldsymbol{\mathrm{e}}} \rangle\right)^{-1}.
\end{align}

\textit{2) Derivation of $q\left( \boldsymbol{\mathrm{\lambda_h}}\right)$ and $q\left( \boldsymbol{\mathrm{\lambda_e}}\right)$}: We first obtain $q\left( \boldsymbol{\mathrm{\lambda_h}}\right)$ by incorporating \eqref{eq:cond_h} and \eqref{eq:cond_lambh} into \eqref{eq:q_x_i} as follows:
\begin{align}\label{eq:q_lambda_h}
q\left(\boldsymbol{\lambda_{\mathrm{h}}}\right) \propto &\exp \left\{ \langle \log p(\boldsymbol{\mathrm{h}} \vert \boldsymbol{\lambda}_{\boldsymbol{\mathrm{h}}}) \rangle_{-\boldsymbol{\lambda_{\mathrm{h}}}}  + \log p(\boldsymbol{\lambda}_{\boldsymbol{\mathrm{h}}}) \right\} \notag \\
\propto &\prod_{i=1}^{D_{\mathrm{U}}D_{\mathrm{B}}} \lambda^{\left(a+1\right)-1}_{\boldsymbol{\mathrm{h}},i} \exp\left\{ -\left( b+ \left[ \boldsymbol{\Sigma_{\mathrm{h}}} \right]_{i,i} + \langle \vert \mathrm{h}_i \vert^2 \rangle  \right) \lambda_{\boldsymbol{\mathrm{h}},i} \right\}, 
\end{align}
which is a product of independent gamma distributions with respect to the same shape parameter $\bar{a}_{\boldsymbol{\mathrm{h}}}=a+1$ and different rates $\bar{b}_{{\boldsymbol{\mathrm{h}}},i}=b+\left[ \boldsymbol{\Sigma_{\mathrm{h}}} \right]_{i,i} + \langle \vert \mathrm{h}_i \vert^2 \rangle  = b + \left[ \boldsymbol{\Sigma_{\mathrm{h}}} + \boldsymbol{\mu_{\mathrm{h}}} \boldsymbol{\mu}^{\mathrm{H}}_{\boldsymbol{\mathrm{h}}} \right]_{i,i} $, where $\left[\bA\right]_{i,i}$ denotes the $i$-th diagonal element of matrix $\bA$. Therefore, the posterior mean of $\lambda_{\bh,i}$ is 
\begin{align}\label{eq:q_lambda_h_mean}
\langle \lambda_{\boldsymbol{\mathrm{h}},i} \rangle =\frac{\bar{a}_{\boldsymbol{\mathrm{h}}}}{\bar{b}_{\boldsymbol{\mathrm{h}},i}}. 
\end{align}
In a similar way, $q\left( \boldsymbol{\mathrm{\lambda_e}}\right)$ is computed by substituting \eqref{eq:cond_e} and \eqref{eq:cond_lambe} into \eqref{eq:q_x_i} as follows: 
\begin{align}\label{eq:q_lambda_e}
q\left(\boldsymbol{\lambda_{\mathrm{e}}}\right) \propto &\exp \left\{ \langle \log p(\boldsymbol{\mathrm{e}} \vert \boldsymbol{\lambda}_{\boldsymbol{\mathrm{e}}}) \rangle_{-\boldsymbol{\lambda_{\mathrm{e}}}}  + \log p(\boldsymbol{\lambda}_{\boldsymbol{\mathrm{e}}}\vert \boldsymbol{\gamma}) \right\} \notag \\
\propto &\prod_{j=1}^{N_{\mathrm{r}}T} \lambda^{\left(\frac{1}{2}-1\right)}_{\boldsymbol{\mathrm{e}},j} \exp \left\{ - \frac{\langle \gamma_j \rangle}{4} \lambda_{\boldsymbol{\mathrm{e}},j} \right\} \notag \\ &\times \exp \left\{ \left( \left[ \boldsymbol{\Sigma_{\mathrm{e}}} \right]_{j,j} + \langle \vert \mathrm{e}_{j} \vert^2 \rangle \right) \lambda^{-1}_{\boldsymbol{\mathrm{e}},j} \right\}.
\end{align}
This implies that $q(\lambda_{\mathbf{e},j})$ follows a generalized inverse Gaussian distribution, where the posterior means of $\lambda_{\mathbf{e},j}$ and $\lambda_{\mathbf{e},j}^{-1}$ are respectively given by \cite{Babacan:2014}
\begin{align}\label{eq:q_lambda_e_mean}
\langle \lambda_{\boldsymbol{\mathrm{e}},j} \rangle &=\frac{2\sqrt{\left[ \boldsymbol{\Sigma_{\mathrm{e}}}+  \boldsymbol{\mu}_{\boldsymbol{\mathrm{e}}} \boldsymbol{\mu}^{\mathrm{H}}_{\boldsymbol{\mathrm{e}}} \right]_{j,j}}}{\sqrt{\langle \gamma_j \rangle}} + \frac{2}{\langle \gamma_j \rangle}, \\ \label{eq:q_lambda_e_inv_mean}
\langle \lambda^{-1}_{\boldsymbol{\mathrm{e}},j} \rangle &= \frac{\sqrt{\langle \gamma_j \rangle}}{2\sqrt{\left[ \boldsymbol{\Sigma_{\mathrm{e}}}+  \boldsymbol{\mu}_{\boldsymbol{\mathrm{e}}} \boldsymbol{\mu}^{\mathrm{H}}_{\boldsymbol{\mathrm{e}}} \right]_{j,j}}}. 
\end{align}

\begin{algorithm}[t]
\caption{Proposed VI-based channel estimator}
\label{algorithm1}
\textbf{Input:} $\boldsymbol{\mathrm{y}}$, $\bPhi$ \\
\textbf{Output:} $\boldsymbol{\mu}_{\mathbf{h}}$ and $\boldsymbol{\mu}_{\mathbf{e}}$
\begin{algorithmic}[1]
    \State Set the hyperparameters $a, b$, $\epsilon_{\mathbf{h}}$ and $\epsilon_{\mathbf{e}}$
    \State Initialize variables:
        \item \quad $\boldsymbol{\mu}_{\bh}=\mathbf{0}\in \mathbb{C}^{D_\mathrm{U}D_\mathrm{B}\times 1}$, $\boldsymbol{\mu}_{\mathbf{e}}=\mathbf{0}\in \mathbb{C}^{N_\mathrm{r}T\times 1}$
        \item \quad $\boldsymbol{\Sigma}_{\bh} = \mathbf{0} \in \mathbb{C}^{D_{\mathrm{U}}D_{\mathrm{B}}\times D_{\mathrm{U}}D_{\mathrm{B}}}$,  $\boldsymbol{\Sigma}_{\mathbf{e}} = \mathbf{0}\in \mathbb{C}^{N_\mathrm{r}T\times N_\mathrm{r}T}$
        \item \quad $\langle \lambda_{\bh, i} \rangle = a/b$, $\forall i$, $\langle \gamma_{i} \rangle = a/b$, $\forall i$, $\langle \beta \rangle = a/b$
        \item \quad $\langle \lambda_{\mathbf{e}, j} \rangle = 1/\langle \gamma_j \rangle$, $\forall j$,  $\langle \lambda_{\mathbf{e}, j}^{-1} \rangle = 1/\langle \lambda_{\mathbf{e}, j} \rangle$, $\forall j$        
    \Repeat
        \State Update $\boldsymbol{\mu}_{\mathbf{h}}$ and $\boldsymbol{\Sigma}_{\bh}$ using \eqref{mean_h}
        \State Update $\boldsymbol{\mu}_{\mathbf{e}}$ and $\boldsymbol{\Sigma}_{\mathbf{e}}$ using \eqref{mean_e}
        \State Update $\langle \lambda_{\bh, i} \rangle$, $\forall i$ using \eqref{eq:q_lambda_h_mean}
        \State Update $\langle \lambda_{\mathbf{e}, j} \rangle$ and $\langle \lambda^{-1}_{\mathbf{e}, j} \rangle$, $\forall j$ using \eqref{eq:q_lambda_e_mean} and \eqref{eq:q_lambda_e_inv_mean}
        \State Update $\langle \gamma_j \rangle$, $\forall j$ using \eqref{eq:gamma_mean}
        \State Update $\langle \beta \rangle$ using \eqref{eq:beta_mean}
    \Until{$\|\Delta\boldsymbol{\mu}_{\mathbf{h}}\| < \epsilon_{\mathbf{h}}$ and $\|\Delta\boldsymbol{\mu}_{\mathbf{e}}\| < \epsilon_{\mathbf{e}}$}
\end{algorithmic}
\end{algorithm}

\textit{3) Derivation of $q\left( \boldsymbol{\gamma} \right)$ and $q\left( \beta \right)$}: To derive $q\left( \boldsymbol{\gamma} \right)$, we plug in \eqref{eq:cond_lambe} and \eqref{eq:gamma} to \eqref{eq:q_x_i} as
\begin{align}\label{eq:q_gamma}
q\left(\boldsymbol{\gamma}\right) \propto &\exp \left\{ \langle \log p( \boldsymbol{\lambda}_{\boldsymbol{\mathrm{e}}} \vert \boldsymbol{\gamma} ) \rangle_{-\boldsymbol{\gamma}}  + \log p(\boldsymbol{\gamma}) \right\} \notag \\
\propto &\prod_{j=1}^{N_{\mathrm{r}}T} \gamma^{\left(a+\frac{3}{2}\right)-1}_{j} \exp\left\{ -\left( b+ \frac{\langle \lambda_{\boldsymbol{\mathrm{e}}, j} \rangle}{4}  \right) \gamma_{j} \right\}.
\end{align}
This corresponds to a product of independent gamma distributions, with the same shape $\bar{a}_\gamma=a+\frac{3}{2}$ and different rates $\bar{b}_{\gamma_j} = b + \frac{\langle \lambda_{\boldsymbol{\mathrm{e}},j} \rangle}{4}$.
The posterior mean of $\gamma_j$ is then obtained as 
\begin{align}\label{eq:gamma_mean}
\langle \gamma_{j} \rangle =\frac{\bar{a}_{\gamma}}{\bar{b}_{\gamma_j}}. 
\end{align}
Finally, using \eqref{eq:cond_y} and \eqref{eq:beta}, $q\left(\beta\right)$ is obtained as
\begin{align}\label{eq:q_beta}
q\left(\beta\right) \propto &\exp \left\{ \langle \log p(\by \vert \boldsymbol{\mathrm{h}}, \boldsymbol{\mathrm{e}}, \beta) \rangle_{-\beta} + \log p(\beta) \right\} \notag \\
\propto & \beta^{\left(a+N_{\mathrm{r}}T\right)-1} \exp\left\{ -\left( b+C_b \right) \beta \right\},
\end{align}
where $C_b = \Vert \by -\boldsymbol{\Phi} \langle\boldsymbol{\mathrm{h}}\rangle -\langle\boldsymbol{\mathrm{e}}\rangle \Vert^2 + \mathrm{tr} [\boldsymbol{\Phi}^{\mathrm{H}}\boldsymbol{\Phi} \boldsymbol{\Sigma}_{\boldsymbol{\mathrm{h}}}] +\mathrm{tr}[\boldsymbol{\Sigma}_{{\boldsymbol{\mathrm{e}}}}]$. This implies that $q(\beta)$ follows a gamma distribution with $\bar{a}_\beta = a + N_{\mathrm{r}}T$ and $\bar{b}_\beta = b + C_b$. Thus, the posterior mean of $\beta$ is
\begin{align}\label{eq:beta_mean}
\langle \beta \rangle =\frac{\bar{a}_{\beta}}{\bar{b}_{\beta}}. 
\end{align}

The proposed VI-based channel estimator is summarized in Algorithm~\ref{algorithm1}. It sequentially updates the posterior means of $\cX$. The algorithm terminates when the iteration-to-iteration changes in $\boldsymbol{\mu}_\bh$ and $\boldsymbol{\mu}_\mathbf{e}$ fall below their respective thresholds $\epsilon_{\mathbf{h}}$ and $\epsilon_{\mathbf{e}}$.
Note that the MMSE estimate of the gNB-UE channel from the proposed channel estimator is given by $\hat{\bar{\bH}}=\mathbf{A}_\mathrm{U}\hat{\mathbf{H}}\mathbf{A}^{\mathrm{H}}_\mathrm{B}$, where $\hat{\bH}$ is reconstructed from $\boldsymbol{\mu}_{\bh}$.


\begin{figure}[!t]
    \centering
    \begin{subfigure}[b]{0.24\textwidth}
        \centering
        \includegraphics[width=\textwidth]{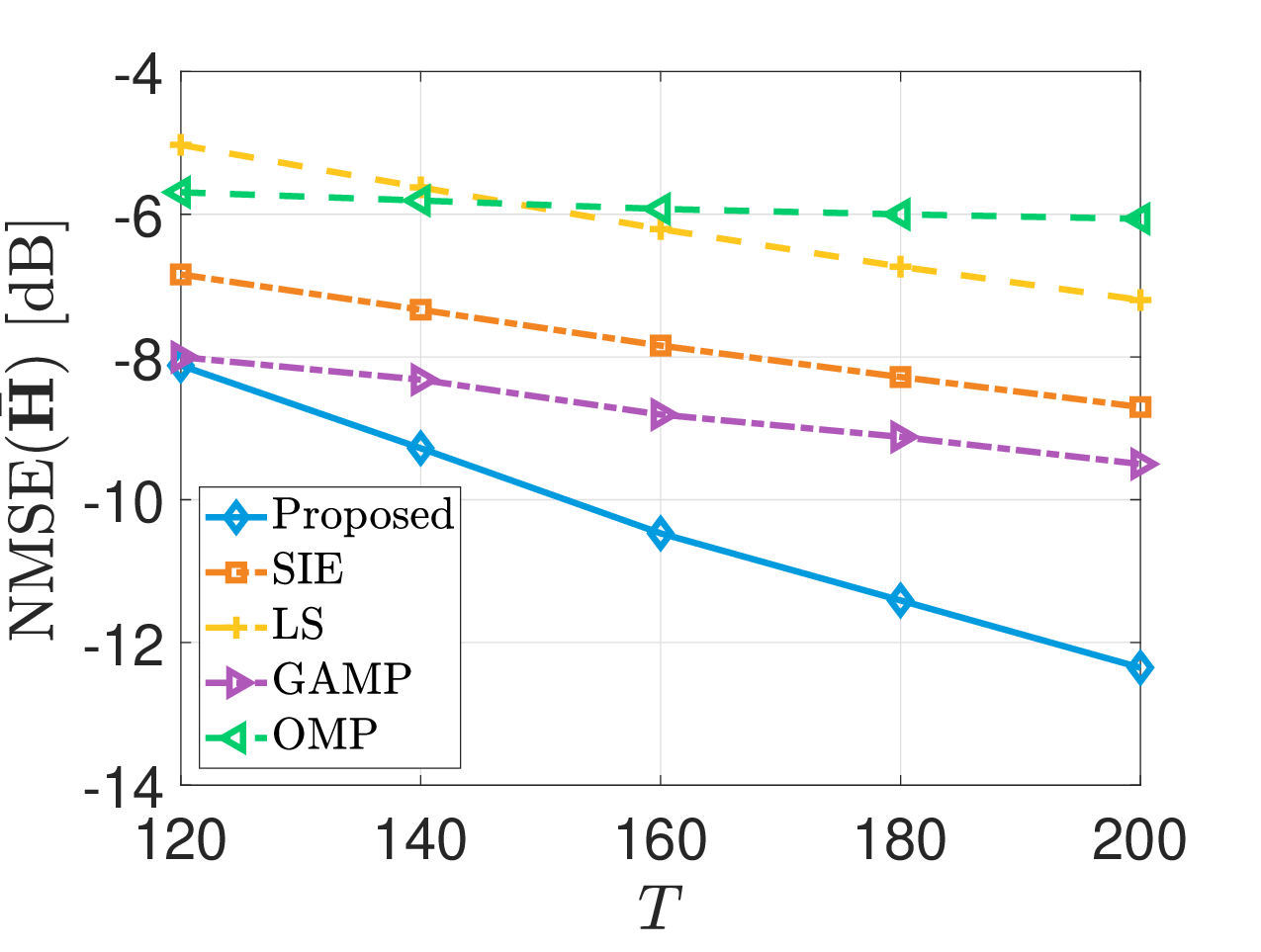}
        \caption{Number of time slots $T$}
        \label{fig:simulResult_time}
    \end{subfigure}
    \hfill
    \begin{subfigure}[b]{0.24\textwidth}
        \centering
        \includegraphics[width=\textwidth]{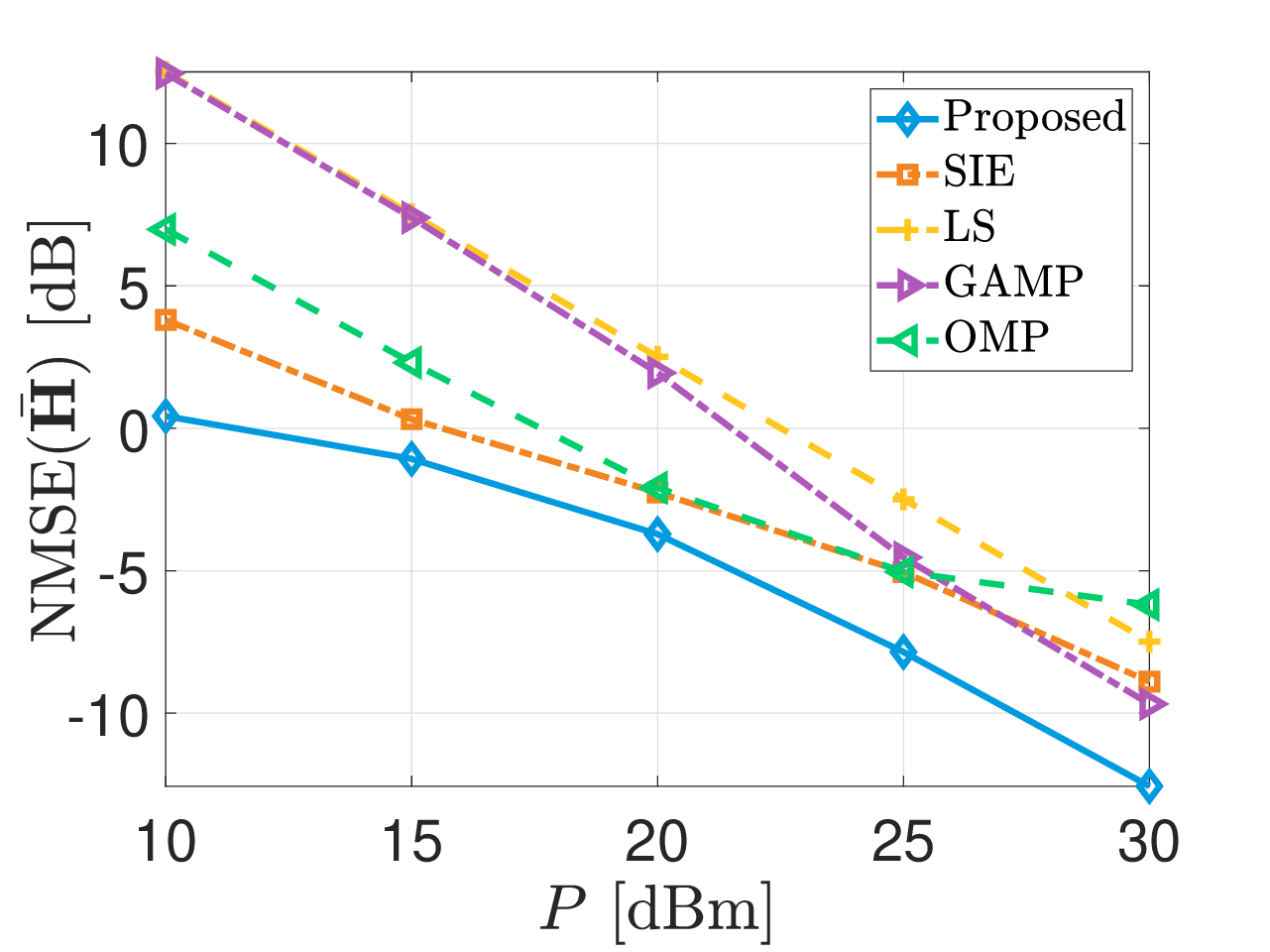}
        \caption{Per-antenna transmit power $P$}
        \label{fig:simulResult_power}
    \end{subfigure}
    \caption{NMSE comparison with respect to $T$ and $P$.}
    \label{fig:nmse_comparison}
\end{figure}

\section{Simulation Results}
We consider a system with $N_{\mathrm{t}}=16$ gNB antennas and $N_{\mathrm{r}}=4$ UE antennas, where the UE is located 50~$\mathrm{m}$ from the gNB. The carrier frequency is set to $f_c = 28~\mathrm{GHz}$, corresponding to the spectrum sweet spot of the mmWave, i.e., n261 band \cite{3gpp.38.104}. The noise spectral density is assumed to be $-174~\mathrm{dBm/Hz}$, with a 100~$\mathrm{MHz}$ bandwidth, which results in the noise variance $\sigma^2=-94~\mathrm{dBm}$. 
The impulsive interference is modeled using a two-component Gaussian mixture model (GMM), a widely adopted approach for capturing its statistical characteristics \cite{dai2017sparse}. The two-component GMM is parameterized by the occurrence probabilities $c_k$ and variance $\sigma_k^2$ for $k \in \left\{1, 2\right\}$, satisfying $c_1 + c_2 = 1, \sigma_2^2 = \eta \sigma_1^2$, and $c_1 > c_2 > 0$. Here, the first component, $k=1$, corresponds to AWGN, while the second component, $k=2$, represents the impulsive interference, and $\eta$ stands for the power ratio between AWGN and impulsive interference. The shape and rate parameters for the gamma distribution in the proposed approach are set to $a=b=10^{-6}$, adhering to the convention of uninformative priors. The convergence thresholds $\epsilon_{\mathbf{h}}$ and $\epsilon_{\mathbf{e}}$ are both set to $10^{-3}$.
Unless otherwise specified, we set $P = 30~\mathrm{dBm}$ as the per-antenna transmit power, along with $T=200$, $c_2=0.1$, and $\eta=10^5$.

To assess the effectiveness of the proposed technique, we compare its performance against the following baselines: least squares (LS), orthogonal matching pursuit (OMP) \cite{cai2011orthogonal}, generalized approximate message passing (GAMP) \cite{rangan2011generalized}, and Student-$t$ based interference estimation (SIE), which models the impulsive interference using the Student-$t$ distribution to capture its non-Gaussian characteristics.
As the performance metric, we adopt the normalized mean squared error (NMSE) defined as
\begin{align}\label{eq:NMSE}
\mathrm{NMSE}(\bar{\bH}) = \mathbb{E}\left[\frac{\Vert \bar{\bH} - \hat{\bar{\bH}} \Vert_{\mathrm{F}}^2}{\Vert \bar{\bH} \Vert_{\mathrm{F}}^2} \right], 
\end{align}
where $\Vert \bA \Vert_{\mathrm{F}}$ denotes the Frobenius norm for a matrix $\bA$.



Fig.~\ref{fig:simulResult_time} shows the NMSE versus the number of time slots $T$. 
When $T$ is sufficiently large, the proposed technique outperforms all baseline approaches, and the performance gap between the proposed technique and baselines increases with $T$ due to accurate interference estimation with sufficient observations. The performance gap between the proposed technique and the SIE indicates that the complex adaptive Laplace prior in our approach more effectively captures the characteristics of impulsive interference.

\begin{figure}[!t]
    \centering
    \begin{subfigure}[b]{0.24\textwidth}
        \centering
        \includegraphics[width=\textwidth]{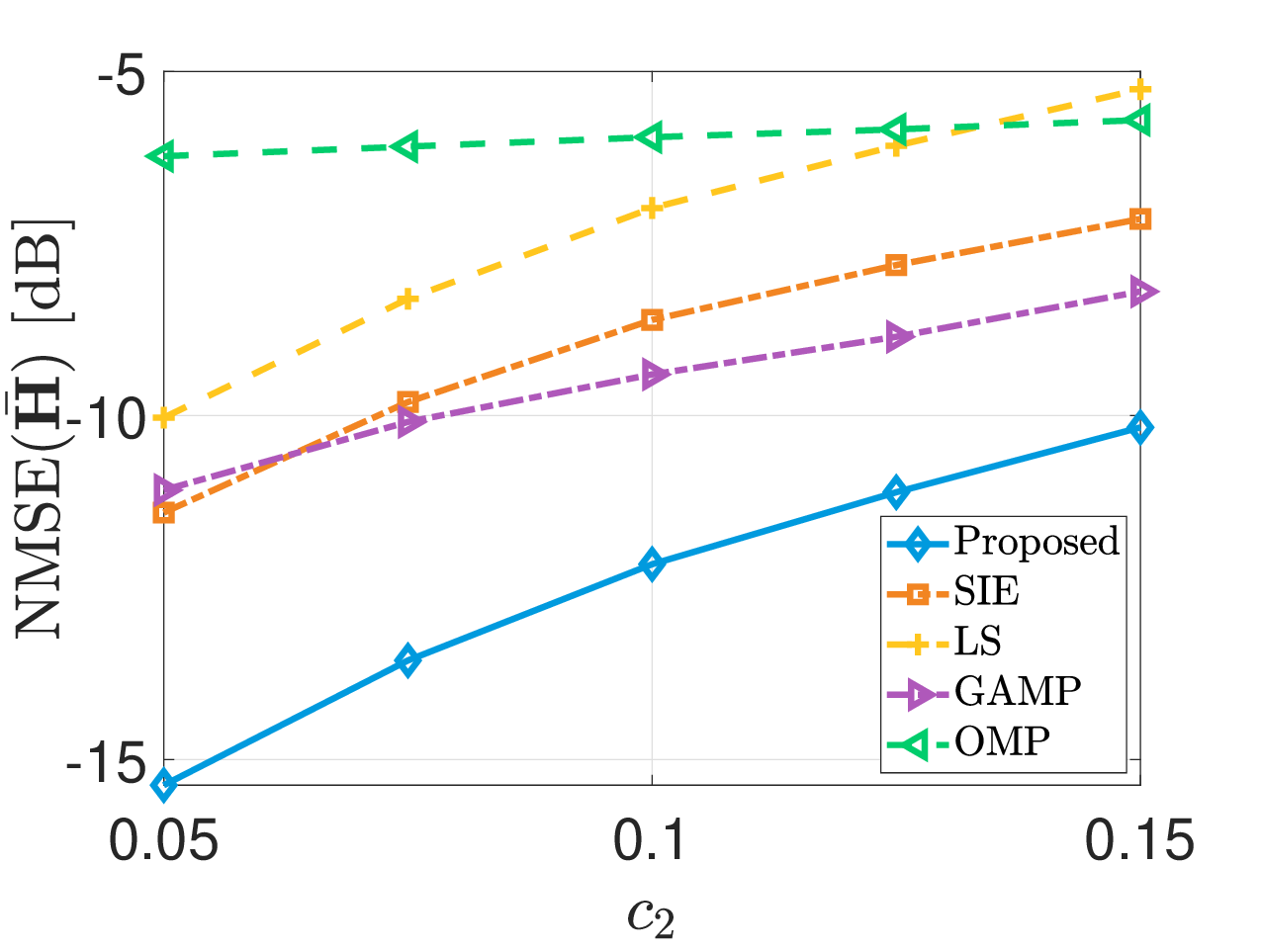}
        \centering
        \subcaption{Occurrence probability of impulsive interference $c_2$}
        \label{fig:simulResult_intProb}
    \end{subfigure}
    \hfill
    \begin{subfigure}[b]{0.24\textwidth}
        \centering
        \includegraphics[width=\textwidth]{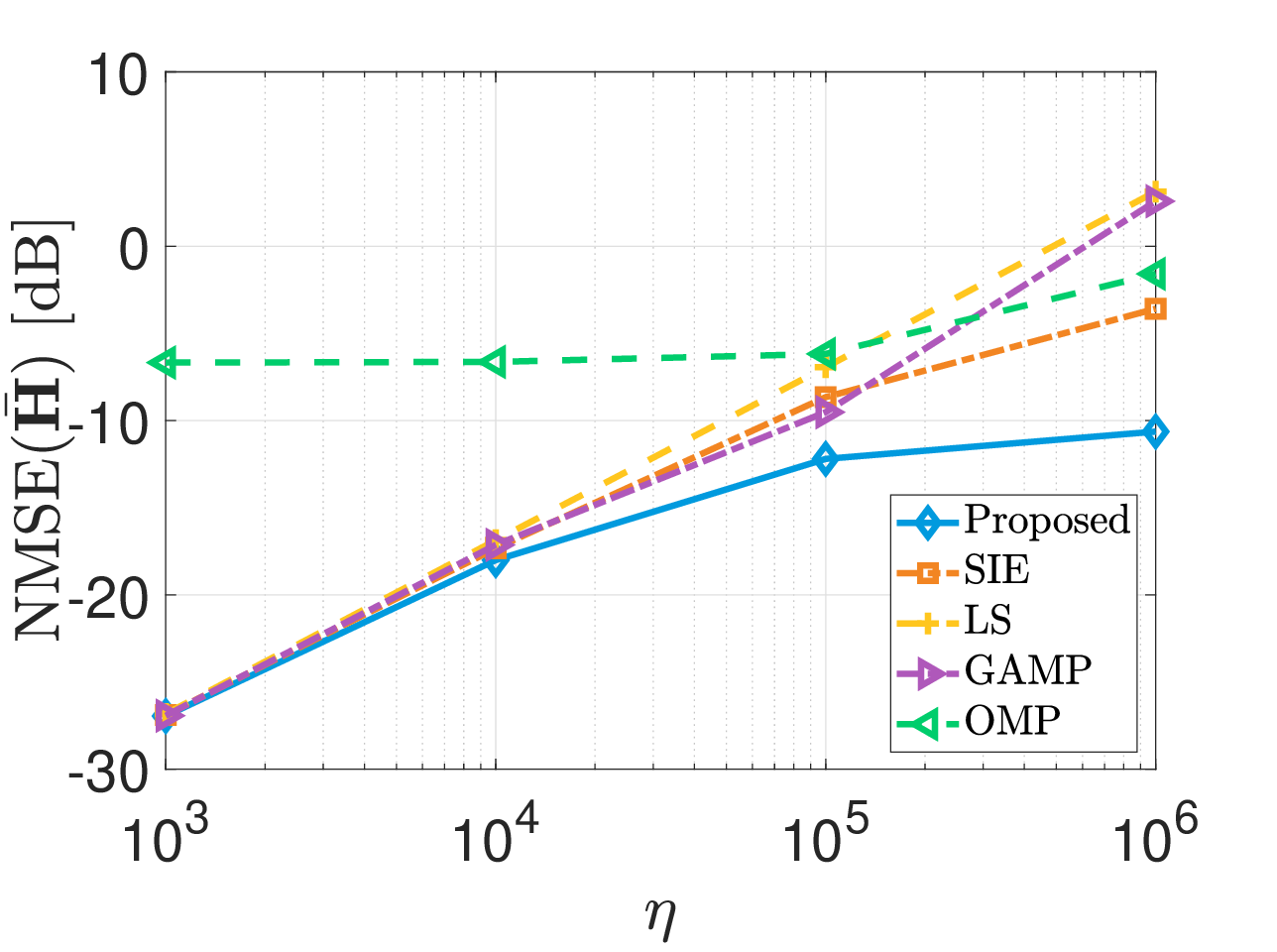}
        \subcaption{Impulsive interference power ratio $\eta$}
        \label{fig:simulResult_intPwr}
    \end{subfigure}
    \caption{NMSE comparison with respect to $c_2$ and $\eta$.}
    \label{fig:nmse_comparison}
\end{figure}

Fig.~\ref{fig:simulResult_power} presents the NMSE comparison with respect to the per-antenna transmit power $P$. 
Increasing $P$ makes the impulsive interference relatively weak, leading to the performance improvements for all baselines. Similar to the results shown in Fig. \ref{fig:simulResult_time}, the proposed technique consistently outperforms all baselines across the entire power range, as a result of the joint estimation of the desired channel and impulsive interference. 

Now, we investigate the NMSE comparisons by varying the characteristics of the impulsive interference, as depicted in Figs.~\ref{fig:simulResult_intProb} and~\ref{fig:simulResult_intPwr}. 
First, Fig. \ref{fig:simulResult_intProb} shows the NMSE comparison versus the occurrence probability of impulsive interference $c_2$.
All approaches show increasing NMSE with $c_2$ since a higher occurrence probability of impulsive interference leads to more severe interference effects. In particular, as $c_2$ increases, its characteristics become less impulsive, making the impulsive interference matrix $\bE$ less sparse. Consequently, the Laplace prior in the proposed technique does not capture the actual interference well, contributing to performance degradation.
Nevertheless, the proposed technique consistently outperforms all baselines, demonstrating that the estimation performance of the desired channel can be significantly improved even in the presence of impulsive interference.

Fig.~\ref{fig:simulResult_intPwr} shows the NMSE comparison according to the impulsive interference power ratio $\eta$. It is observed that the performance of the proposed technique is comparable to that of the baseline approaches when the impulsive interference is relatively weak and similar to background noise.
However, as $\eta$ increases, the NMSE of the baselines gradually degrades, whereas the proposed technique not only maintains the lowest NMSE but also exhibits a slower degradation, showing its robust performance even under strong impulsive interference.

Lastly, we also compare the computational complexities as $\mathcal{O}\left(I_{\mathrm{p}} N_{\mathrm{r}}^3 T^3\right)$ for the proposed technique, $\mathcal{O}\left(I_{\mathrm{s}} N_{\mathrm{r}}^3 T^3\right)$ for SIE, $\mathcal{O}\left(I_{\mathrm{g}}  N_{\mathrm{r}} T D_{\mathrm{U}} D_{\mathrm{B}}\right)$ for GAMP, $\mathcal{O}\left( K N_{\mathrm{r}} T D_{\mathrm{U}} D_{\mathrm{B}}\right)$ for OMP, and $\mathcal{O}\left(N_{\mathrm{t}} N_{\mathrm{r}} T\right)$ for LS, where $I_{\mathrm{p}}, I_{\mathrm{s}},$ and $I_{\mathrm{g}}$ are the iteration counts for the proposed technique, SIE, and GAMP, respectively, and $K$ stands for the sparsity level. We note that the proposed technique maintains manageable complexity while achieving superior estimation performance \cite{gadgetversus_exynos2500}.



\section{Conclusions}
\label{sec6}
In this paper, we addressed the downlink channel estimation in mmWave MIMO systems under impulsive interference. To tackle this issue, we proposed a VI-based channel estimator leveraging the SBL framework. Our approach effectively captured the sparsity of the mmWave channel and the sporadic nature of impulsive interference by adopting Student-$t$ and complex adaptive Laplace priors. Simulation results demonstrated that the proposed technique consistently outperforms baselines in severe impulsive interference scenarios. The findings highlight the robustness of our approach in achieving accurate channel estimation under impulsive interference.



\bibliographystyle{IEEEtran}
\bibliography{refs_all}

\vfill
	
\end{document}